\RequirePackage{fix-cm}
\documentclass[smallextended]{svjour3}       
\smartqed  
\usepackage{amssymb}
\usepackage{amsmath}
\usepackage{color}
\usepackage{graphicx}
\usepackage[justification=centering]{caption}
\usepackage{multirow}
\usepackage{marvosym}
\begin{document}

\title{Deterministic Joint Remote Preparation of a Four-qubit Cluster-type State via GHZ states
} \subtitle{}


\author{Hai-bin Wang       \and
        Xiao-Yan Zhou       \and
        Xing-xing An    \and
        Meng-Meng Cui    \and
        De-sheng Fu     
}


\institute{
		  H.-B. Wang (\Letter)    
		  \at Jiangsu Engineering Center of Network Monitoring, Nanjing University of Information Science \& Technology, Nanjing 210044, P. R. China
		  \\
		  \email{whb787716@163.com}
		  \and
           H.-B. Wang \and
           M.-M. Cui  \and
           D.-S. Fu     
           \at School of Computer and Software, Nanjing University of Information Science \& Technology, Nanjing 210044, P. R. China
           \\
           \email{whb787716@163.com}
           \and
           X.-Y. Zhou
           \at Jiangsu Key Laboratory of Meteorological Observation and Information Processing, Nanjing 210044, P. R. China
           \and
           X.-Y. Zhou \and
           X.-X. An
           \at School of Electronic and Information Engineering, Nanjing University of Information Science and Technology, Nanjing 210044, P. R. China
}

\date{Received: August 2, 2015 / Accepted: date}

\maketitle

\begin{abstract}
A scheme for the deterministic joint remote preparation of a four-qubit cluster-type state using only two Greenberger-Horne-Zeilinger (GHZ) states as quantum channels is presented. In this scheme, the first sender performs a two-qubit projective measurement according to the real coefficient of the desired state. Then, the other sender utilizes the measurement result and the complex coefficient to perform another projective measurement. To obtain the desired state, the receiver applies appropriate unitary operations to his/her own two qubits and two \emph{CNOT} operations to the two ancillary ones. Most interestingly, our scheme can achieve unit success probability, i.e., $P_{suc}$=1. Furthermore, comparison reveals that the efficiency is higher than that of most other analogous schemes.

\end{abstract}

\section{Introduction}
\label{intro1}
With the rapid development of quantum information technology in recent
decades, quantum entanglement has begun to play a very critical role
as a physical resource. Many quantum information protocols have been
developed based on quantum mechanics principles, including quantum key
distribution [1,2], quantum secure direct communication [3,4], quantum teleportation (QT) [5,6], and quantum private
comparison [7-9].  Especially in QT, the preparer can transfer an
unknown quantum state to a remote receiver using classical information
and quantum resources. A major characteristic of QT is that the
preparer knows nothing about the state. In the last decade, Lo [10],
Pati [11] and Bennett et al. [12] reported a new quantum communication
scheme that uses classical communication and a previously shared
entangled resource to remotely prepare a quantum state. This scheme is
called remote state preparation (RSP). Similar to QT, in RSP, the
preparer can exploit the nonlocal correlation of the quantum entangled
state that is shared in advance to prepare the remote desired
state. The main difference is that in RSP, the preparer must know all
the state information, while in QT, the preparer knows nothing about
the state. To date, because RSP has potential value in quantum communication, it has been widely studied, and many RSP protocols have been proposed [13,14].

The early RSP protocols focused on the case of one preparer and one receiver, in which the preparer knows all the desired state information. However, it is unreliable that one one preparer holds all the information, especially this information is very important and highly sensitive. To overcome this issue, joint remote state preparation (JRSP) was developed. In 2007, Xia et al. [15] proposed the novel JRSP protocol, which realized the multiparty remote preparation of an arbitrary one-qubit state. Since then, many JRSP schemes have been presented, including the preparation of one-qubit [16-18], two-qubit [19-21], and three-qubit states [22-24].

Recently, some researchers have begun to focus on more complicated
four-qubit cases using a variety of methods. For instance, in 2011,
Zhan et al. [25] investigated JRSP for a four-qubit cluster-type state
with six Einstein-Podolsky-Rosen(EPR) pairs and two six-qubit entangled states (ZHM11 for
short), and subsequently, An et al. [26] proposed a different scheme
that only requires the projective measurement of two qubits and fewer
classical bits (ABD11). Moreover, some researchers have attempted to
prepare cluster-type states using partially entangled states as
quantum channels. In 2012, Wang et al. [27] addressed this issue using
two quaternate partially entangled states as quantum channels
(WY12). In 2013, Wang et al. [28] exploited a new and feasible scheme
to perform JRSP of four-qubit cluster-type states based on tripartite
non-maximal entanglements and positive operator-valued measurement
(WY13). Additionally, Hou [29] proposed a scheme for JRSP of four-qubit
cluster-type states using only two-qubit entangled states as the
quantum channel (H13), which he then extended to the multiparty
case. Unfortunately, these protocols are probabilistic and therefore
cannot be realized with unit success probability. To solve this
problem, by tactically constructing two sets of projective measurement
bases, i.e., the real-coefficient measurement basis and the
complex-coefficient measurement basis, a new deterministic JRSP scheme
with GHZ states is proposed. In this scheme, the first sender performs
a two-qubit projective measurement based on the real coefficient of
the desired state. Then, the other sender performs another projective
measurement utilizing the measurement result and the complex
coefficient. To obtain the desired state, the receiver applies
appropriate unitary operations to his/her own two qubits and two
\emph{CNOT} operations to the two ancillary ones. This JRSP scheme can be successfully realized with unit success probability.

The paper is organized as follows. In the next section, we propose a
new scheme for joint remotely preparing a four-qubit cluster-type
state and provide a detailed description. We analyze the correctness of the protocol in the third section. Finally, a concise summary is given in the last section.

\section{Deterministic JRSP of a four-qubit cluster-type state via GHZ states}
\label{sec:2}
Without loss of generality, suppose there are three parties--two senders Alice and Bob and one receiver Charlie--and the senders cooperate to prepare a four-qubit cluster-type state for the receiver,

\begin{equation}\label{eq:schemeP}
|\phi \rangle  = a|0000\rangle  + b{e^{i{\theta _1}}}|0011\rangle {\kern 1pt}  + c{e^{i{\theta _2}}}|1100\rangle  + d{e^{i{\theta _3}}}|1111\rangle,
\end{equation}
where $a$, $b$, $c$, $d$ and ${\theta _i}$ (i=1,2,3) are real and meet the
normalized condition ${a^2} + {b^2} + {c^2} + {d^2} =
1$. Additionally, Alice and Bob have partial knowledge of the desired
state $\left| \phi  \right\rangle$; that is, Alice only knows the real
coefficients a, b, c and d, while Bob knows the complex coefficient ${\theta _i}$. In this scenario, neither Alice nor Bob can individually help Charlie to reconstruct the desired state.

\begin{figure}[h]
\begin{center}
\includegraphics[width=8cm]{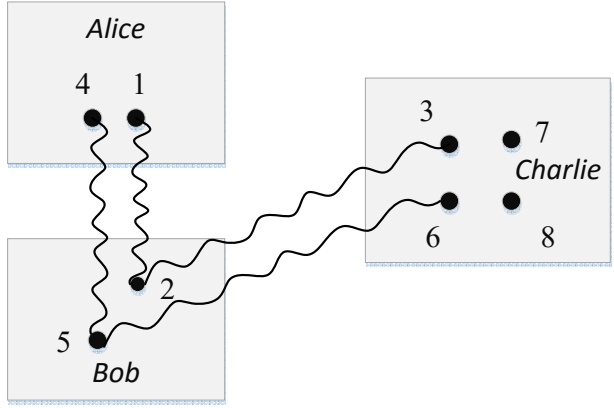}
\caption{{\bf The schematic distribution of qubits used to establish the correlation among the participants in our scheme.}}
\label{fig1}
\end{center}
\end{figure}

Suppose the quantum channel shared by Alice, Bob and Charlie consists of two GHZ states,
\begin{equation}\label{eq:eq1}
\left| Q \right\rangle  = \frac{1}{{\sqrt 2 }}(|000\rangle  + |111\rangle {)_{123}} \otimes \frac{1}{{\sqrt 2 }}(|000\rangle  + |111\rangle {)_{456}},
\end{equation}
where qubits 1 and 4 belong to Alice, qubits 2 and 5 belong to Bob,
and qubits 3 and 6 belong to Charlie (see Fig.~\ref{fig1}).

To help Charlie remotely prepare the desired state, the detailed four-step protocol can be described as below:

\textbf{Step 1} Alice performs a two-qubit projective measurement on
her own qubits 1 and 4. Here, the measurement bases are a set of
mutually orthogonal basis vectors constructed using the real part $\left\{ {a,b,c,d} \right\}$ as below,
\begin{equation}\label{eq:schemeP}
\left( {\begin{array}{*{20}{c}}
{|{u_0}{\rangle _{14}}}\\
{|{u_1}{\rangle _{14}}}\\
{|{u_2}{\rangle _{14}}}\\
{|{u_3}{\rangle _{14}}}
\end{array}} \right) = U\left( {a,b,c,d} \right)\left( {\begin{array}{*{20}{c}}
{|00{\rangle _{14}}}\\
{|01{\rangle _{14}}}\\
{|10{\rangle _{14}}}\\
{|11{\rangle _{14}}}
\end{array}} \right),
\end{equation}
where
\begin{equation}\label{eq:schemeP}
U\left( {a,b,c,d} \right) = \left( {\begin{array}{*{20}{c}}
a&b&c&d\\
b&{ - a}&d&{ - c}\\
c&{ - d}&{ - a}&b\\
d&c&{ - b}&{ - a}
\end{array}} \right).
\end{equation}
Then, Eq.~(\ref{eq:eq1}) can be rewritten as
\begin{equation}\label{eq:schemeP}
\left| Q \right\rangle  = \frac{1}{2}\sum\limits_{m = 0}^3 {{{\left| {{u_m}} \right\rangle }_{14}}{{\left| {{L_m}} \right\rangle }_{2536}}},
\end{equation}
where
\begin{equation}\label{eq:schemeP}
{\left| {{L_0}} \right\rangle _{2536}} = (a|0000\rangle  + b|0101\rangle {\kern 1pt}  + c|1010\rangle  + d|1111\rangle {)_{2536}},
\end{equation}
\begin{equation}\label{eq:schemeP}
{\left| {{L_1}} \right\rangle _{2536}} = (b|0000\rangle  - a|0101\rangle {\kern 1pt} {\kern 1pt}  + d|1010\rangle  - c|1111\rangle {)_{2536}},
\end{equation}
\begin{equation}\label{eq:schemeP}
{\left| {{L_2}} \right\rangle _{2536}} = (c|0000\rangle  - d|0101\rangle {\kern 1pt} {\kern 1pt}  - a|1010\rangle  + b|1111\rangle {)_{2536}},
\end{equation}
\begin{equation}\label{eq:schemeP}
{\left| {{L_3}} \right\rangle _{2536}} = (d|0000\rangle  + c|0101\rangle  - b|1010\rangle  - a|1111\rangle {)_{2536}}.
\end{equation}
Alice transmits her measurement result to both Charlie and Bob.

\textbf{Step 2} Using $m$ and the complex part ${\theta _i}$, Bob constructs a set of mutually orthogonal basis vectors of qubits 2 and 5:
\begin{equation}\label{eq:schemeP}
\left( {\begin{array}{*{20}{c}}
{|{v_0}\rangle }\\
{|{v_1}\rangle }\\
{|{v_2}\rangle }\\
{|{v_3}\rangle }
\end{array}} \right) = {G^{(m)}}\left( {\begin{array}{*{20}{c}}
{|00\rangle }\\
{|01\rangle }\\
{|10\rangle }\\
{|11\rangle }
\end{array}} \right),
\end{equation}
where
\begin{equation}\label{eq:schemeP}
{G^{(0)}} = \frac{1}{2}\left( {\begin{array}{*{20}{c}}
1&{{e^{ - i{\theta _1}}}}&{{e^{ - i{\theta _2}}}}&{{e^{ - i{\theta _3}}}}\\
1&{ - {e^{ - i{\theta _1}}}}&{{e^{ - i{\theta _2}}}}&{ - {e^{ - i{\theta _3}}}}\\
1&{ - {e^{ - i{\theta _1}}}}&{ - {e^{ - i{\theta _2}}}}&{{e^{ - i{\theta _3}}}}\\
1&{{e^{ - i{\theta _1}}}}&{ - {e^{ - i{\theta _2}}}}&{ - {e^{ - i{\theta _3}}}}
\end{array}} \right){\kern 1pt} {\kern 1pt} {\kern 1pt},
\end{equation}
\begin{equation}\label{eq:schemeP}
{G^{(1)}} = \frac{1}{2}\left( {\begin{array}{*{20}{c}}
{{e^{ - i{\theta _1}}}}&1&{{e^{ - i{\theta _3}}}}&{{e^{ - i{\theta _2}}}}\\
{{e^{ - i{\theta _1}}}}&{ - 1}&{{e^{ - i{\theta _3}}}}&{ - {e^{ - i{\theta _2}}}}\\
{{e^{ - i{\theta _1}}}}&{ - 1}&{ - {e^{ - i{\theta _3}}}}&{{e^{ - i{\theta _2}}}}\\
{{e^{ - i{\theta _1}}}}&1&{ - {e^{ - i{\theta _3}}}}&{ - {e^{ - i{\theta _2}}}}
\end{array}} \right),
\end{equation}
\begin{equation}\label{eq:schemeP}
{G^{(2)}} = \frac{1}{2}\left( {\begin{array}{*{20}{c}}
{{e^{ - i{\theta _2}}}}&{{e^{ - i{\theta _3}}}}&1&{{e^{ - i{\theta _1}}}}\\
{{e^{ - i{\theta _2}}}}&{ - {e^{ - i{\theta _3}}}}&1&{ - {e^{ - i{\theta _1}}}}\\
{{e^{ - i{\theta _2}}}}&{ - {e^{ - i{\theta _3}}}}&{ - 1}&{{e^{ - i{\theta _1}}}}\\
{{e^{ - i{\theta _2}}}}&{{e^{ - i{\theta _3}}}}&{ - 1}&{ - {e^{ - i{\theta _1}}}}
\end{array}} \right){\kern 1pt} {\kern 1pt},
\end{equation}
\begin{equation}\label{eq:schemeP}
{G^{(3)}} = \frac{1}{2}\left( {\begin{array}{*{20}{c}}
{{e^{ - i{\theta _3}}}}&{{e^{ - i{\theta _2}}}}&{{e^{ - i{\theta _1}}}}&1\\
{{e^{ - i{\theta _3}}}}&{ - {e^{ - i{\theta _2}}}}&{{e^{ - i{\theta _1}}}}&{ - 1}\\
{{e^{ - i{\theta _3}}}}&{ - {e^{ - i{\theta _2}}}}&{ - {e^{ - i{\theta _1}}}}&1\\
{{e^{ - i{\theta _3}}}}&{{e^{ - i{\theta _2}}}}&{ - {e^{ - i{\theta _1}}}}&{ - 1}
\end{array}} \right).
\end{equation}
In terms of these measurement bases, ${\left| {{L_m}} \right\rangle _{2536}}$ can be expressed in the form
\begin{equation}\label{eq:schemeP}
{\left| {{L_m}} \right\rangle _{2536}} = \sum\limits_{n = 1}^3 {{{\left| {{v_n}} \right\rangle }_{25}}{{\left| {{D_{mn}}} \right\rangle }_{36}}};
\end{equation}
here, ${\left| {{D_{mn}}} \right\rangle _{36}}$ represents the final
states of qubits 3 and 6. Bob performs the projection measurement on
qubits 2 and 5 and sends the result $n\left( {n = 0,1,2,3} \right)$ to Charlie.

\textbf{Step 3} According to the measurement results $m$ and $n$, Charlie utilizes unitary operators ${R_{mn}}$ (shown in Table~\ref{table1}) to transfer ${\left| {{D_{mn}}} \right\rangle _{36}}$ to the below state,
\begin{equation}\label{eq:schemeP}
{\left| T \right\rangle _{36}} = (a|00\rangle  + b{e^{i{\theta _1}}}|01\rangle  + c{e^{i{\theta _2}}}|10\rangle  + d{e^{i{\theta _3}}}|11\rangle {)_{36}}.
\end{equation}

\label{table1}
\begin{table}[h]\normalsize
\caption{
{\bf The unitary operation of qubits 3 and 6.}}
\renewcommand\arraystretch{1.5}\begin{tabular}{p{2cm}p{2cm}p{7cm}p{2cm}}
\hline {\bf $m$ }& {\bf $n$ }&{\bf ${\left| {{D_{mn}}} \right\rangle _{36}}$}& {\bf ${R_{mn}}$ } \\
\hline
\multirow{4}{1cm}{0}
    & 0 & $ a|00\rangle  + b{e^{i{\theta _1}}}|01\rangle  + c{e^{i{\theta _2}}}|10\rangle  + d{e^{i{\theta _3}}}|11\rangle$ & ${I_3} \otimes {I_6}$ \\
    & 1 & $a|00\rangle  - b{e^{i{\theta _1}}}|01\rangle  + c{e^{i{\theta _2}}}|10\rangle  - d{e^{i{\theta _3}}}|11\rangle$ & ${I_3} \otimes {Z_6}$ \\
    & 2 & $a|00\rangle  - b{e^{i{\theta _1}}}|01\rangle  - c{e^{i{\theta _2}}}|10\rangle  + d{e^{i{\theta _3}}}|11\rangle$ & ${Z_3} \otimes {Z_6}$ \\
    & 3 & $a|00\rangle  + b{e^{i{\theta _1}}}|01\rangle  - c{e^{i{\theta _2}}}|10\rangle  - d{e^{i{\theta _3}}}|11\rangle$ & ${Z_3} \otimes {I_6}$ \\
\hline
\multirow{4}{1cm}{1}
    & 0 & $b{e^{i{\theta _1}}}|00\rangle  - a|01\rangle  + d{e^{i{\theta _3}}}|10\rangle  - c{e^{i{\theta _2}}}|11\rangle$ & ${I_3} \otimes {Z_6}{X_6}$ \\
    & 1 & $b{e^{i{\theta _1}}}|00\rangle  + a|01\rangle  + d{e^{i{\theta _3}}}|10\rangle  + c{e^{i{\theta _2}}}|11\rangle$ & ${I_3} \otimes {X_6}$ \\
    & 2 & $b{e^{i{\theta _1}}}|00\rangle  + a|01\rangle  - d{e^{i{\theta _3}}}|10\rangle  - c{e^{i{\theta _2}}}|11\rangle$ & ${Z_3} \otimes {X_6}$ \\
    & 3 & $b{e^{i{\theta _1}}}|00\rangle  - a|01\rangle  - d{e^{i{\theta _3}}}|10\rangle  + c{e^{i{\theta _2}}}|11\rangle$ & ${Z_3} \otimes {Z_6}{X_6}$ \\
\hline
\multirow{4}{1cm}{2}
    & 0 & $c{e^{i{\theta _2}}}|00\rangle  - d{e^{i{\theta _3}}}|01\rangle  - a|10\rangle  + b{e^{i{\theta _1}}}|11\rangle$ & ${Z_3}{X_3} \otimes {Z_6}$ \\
    & 1 & $c{e^{i{\theta _2}}}|00\rangle  + d{e^{i{\theta _3}}}|01\rangle  - a|10\rangle  - b{e^{i{\theta _1}}}|11\rangle$ & ${Z_3}{X_3} \otimes {I_6}$ \\
    & 2 & $c{e^{i{\theta _2}}}|00\rangle  + d{e^{i{\theta _3}}}|01\rangle  + a|10\rangle  + b{e^{i{\theta _1}}}|11\rangle$ & ${X_3} \otimes {I_6}$ \\
    & 3 & $c{e^{i{\theta _2}}}|00\rangle  - d{e^{i{\theta _3}}}|01\rangle  + a|10\rangle  - b{e^{i{\theta _1}}}|11\rangle$ & ${X_3} \otimes {Z_6}$ \\
\hline
\multirow{4}{1cm}{3}
    & 0 & $d{e^{i{\theta _3}}}|00\rangle  + c{e^{i{\theta _2}}}|01\rangle  - b{e^{i{\theta _1}}}|10\rangle  - a|11\rangle$ & ${Z_3}{X_3} \otimes {X_6}$ \\
    & 1 & $d{e^{i{\theta _3}}}|00\rangle  - c{e^{i{\theta _2}}}|01\rangle  - b{e^{i{\theta _1}}}|10\rangle  + a|11\rangle$ & ${Z_3}{X_3} \otimes {Z_6}{X_6}$ \\
    & 2 & $d{e^{i{\theta _3}}}|00\rangle  - c{e^{i{\theta _2}}}|01\rangle  + b{e^{i{\theta _1}}}|10\rangle  - a|11\rangle$ & ${X_3} \otimes {Z_6}{X_6}$ \\
    & 3 & $d{e^{i{\theta _3}}}|00\rangle  + c{e^{i{\theta _2}}}|01\rangle  + b{e^{i{\theta _1}}}|10\rangle  + a|11\rangle$ & ${X_3} \otimes {X_6}$ \\
\hline
\end{tabular}
\end{table}

\textbf{Step 4} Charlie prepares two ancillary qubits 7 and 8, denoted
as ${\left| {00} \right\rangle _{78}}$, and then applies two
controlled-NOT gates $CNO{T_{3,7}}$ and $CNO{T_{6,8}}$ to qubits 3, 6,
7 and 8. Here, qubits 3 and 6 are the control qubits and qubits 7 and
8 are the target ones. As a result, he obtains the desired four-qubit cluster-type states as below,
\begin{equation}\label{eq:schemeP}
\begin{array}{l}
{CNOT}_{36}{CNOT}_{78}{\left| T \right\rangle _{36}}{\left| {00} \right\rangle _{78}}\\
{\kern 8pt} = {CNOT}_{36}{CNOT}_{78}(a|00\rangle  + b{e^{i{\theta _1}}}|01\rangle  + c{e^{i{\theta _2}}}|10\rangle  + d{e^{i{\theta _3}}}|11\rangle {)_{36}}{\left| {00} \right\rangle _{78}}\\
{\kern 8pt} = a|0000\rangle  + b{e^{i{\theta _1}}}|0011\rangle  + c{e^{i{\theta _2}}}|1100\rangle  + d{e^{i{\theta _3}}}|1111\rangle
\end{array}.
\end{equation}

\begin{figure}[h]
\begin{center}
\includegraphics[width=13cm]{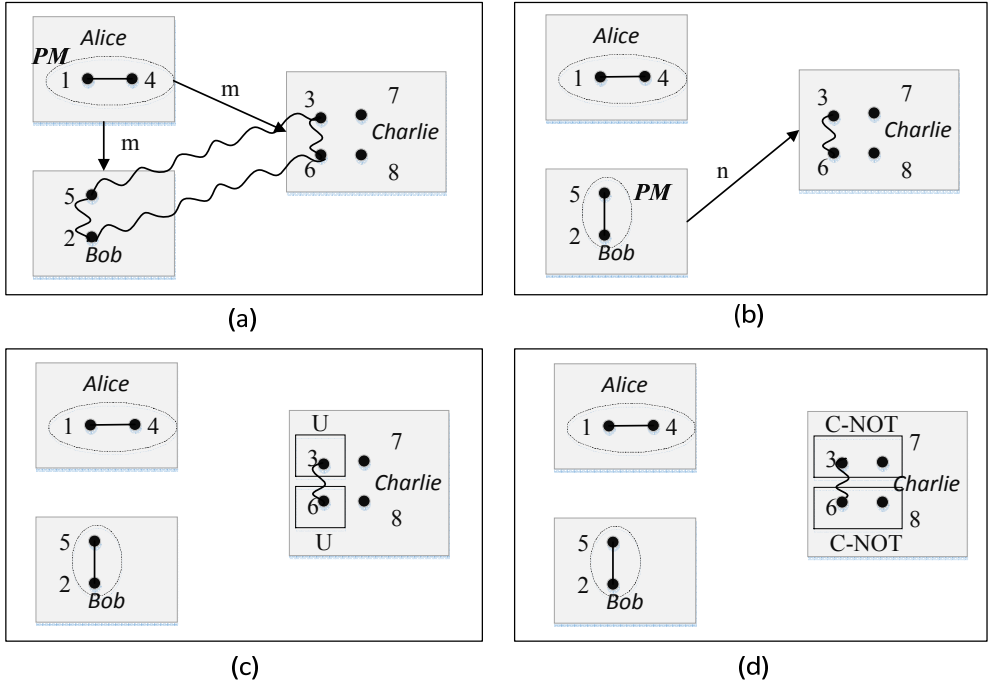}
\caption{{\bf The procedure of our scheme for the deterministic joint
    remote preparation of a four-qubit cluster-type entangled state
    with three parties.} (a) Alice performs a two-qubit projective
  measurement on her own qubits 1 and 4 and sends the measurement
  result m to both Bob and Charlie. (b) Bob performs a projection
  measurement on qubits 2 and 5 and sends the result n to Charlie. (c)
  According to measurement results m and n, Charlie applies unitary
  operations to qubits 3 and 6. (d) Charlie prepares two ancillary
  qubits 7 and 8 and then applies two controlled-NOT gates to qubits 3, 6, 7 and 8 to obtain the desired state.}
\label{fig2}
\end{center}
\end{figure}

The whole process of our four-step JRSP protocol is summarized in Fig.~\ref{fig2}.

\section{Correctness and efficiency analysis}
\label{sec:3}
\subsection{Correctness analysis}
\label{sec:3.1}
Assuming Alice receives the measurement result $m = 1$, Bob chooses the measurement basis according to the strategy of step 2
\begin{equation}\label{eq:schemeP}
\left( {\begin{array}{*{20}{c}}
{|{v_0}\rangle }\\
{|{v_1}\rangle }\\
{|{v_2}\rangle }\\
{|{v_3}\rangle }
\end{array}} \right) = \frac{1}{2}\left( {\begin{array}{*{20}{c}}
{{e^{ - i{\theta _1}}}}&1&{{e^{ - i{\theta _3}}}}&{{e^{ - i{\theta _2}}}}\\
{{e^{ - i{\theta _1}}}}&{ - 1}&{{e^{ - i{\theta _3}}}}&{ - {e^{ - i{\theta _2}}}}\\
{{e^{ - i{\theta _1}}}}&{ - 1}&{ - {e^{ - i{\theta _3}}}}&{{e^{ - i{\theta _2}}}}\\
{{e^{ - i{\theta _1}}}}&1&{ - {e^{ - i{\theta _3}}}}&{ - {e^{ - i{\theta _2}}}}
\end{array}} \right)\left( {\begin{array}{*{20}{c}}
{|00\rangle }\\
{|01\rangle }\\
{|10\rangle }\\
{|11\rangle }
\end{array}} \right).
\end{equation}
We can rewrite ${\left| {{L_1}} \right\rangle _{2536}}$ as
\begin{equation}\label{eq:schemeP}
\begin{array}{l}
{\left| {{L_1}} \right\rangle _{2536}} = \frac{1}{2}\sum\limits_{n = 1}^3 {{{\left| {{v_n}} \right\rangle }_{25}}{{\left| {{D_{1n}}} \right\rangle }_{36}}} \\
{\kern 20pt}  = \frac{1}{2}{\left| {{v_0}} \right\rangle _{25}}{\left( {b{e^{i{\theta _1}}}|00\rangle  - a|01\rangle  + d{e^{i{\theta _3}}}|10\rangle  - c{e^{i{\theta _2}}}|11\rangle } \right)_{36}}\\
{\kern 24pt} + \frac{1}{2}{\left| {{v_1}} \right\rangle _{25}}{\left( {b{e^{i{\theta _1}}}|00\rangle  + a|01\rangle  + d{e^{i{\theta _3}}}|10\rangle  + c{e^{i{\theta _2}}}|11\rangle } \right)_{36}}\\
{\kern 24pt} + \frac{1}{2}{\left| {{v_2}} \right\rangle _{25}}{\left( {b{e^{i{\theta _1}}}|00\rangle  + a|01\rangle  - d{e^{i{\theta _3}}}|10\rangle  - c{e^{i{\theta _2}}}|11\rangle } \right)_{36}}\\
{\kern 24pt} + \frac{1}{2}{\left| {{v_3}} \right\rangle _{25}}{\left( {b{e^{i{\theta _1}}}|00\rangle  - a|01\rangle  - d{e^{i{\theta _3}}}|10\rangle  + c{e^{i{\theta _2}}}|11\rangle } \right)_{36}}
\end{array}.
\end{equation}
When $n = 0$, qubits 3 and 6 will collapse into $b{e^{i{\theta
      _0}}}|00\rangle  - a|01\rangle  + d{e^{i{\theta _3}}}|10\rangle
- c{e^{i{\theta _2}}}|11\rangle$, and the receiver can achieve the
state ${\left| T \right\rangle _{36}} = (a|00\rangle  + b{e^{i{\theta
      _0}}}|01\rangle  + c{e^{i{\theta _2}}}|10\rangle  +
d{e^{i{\theta _3}}}|11\rangle {)_{36}}$ with the identity operator
${I_3} \otimes {Z_6}{X_6}$. According to Step 4, Charlie introduces
the ancillary qubits 7 and 8 into the state ${\left| {00}
  \right\rangle _{78}}$ and then applies two controlled-NOT gates to
obtain the desired state. Similarly, when $n = 1,2,3$, Charlie can
choose the operators ${I_3} \otimes {X_6}$, ${Z_3} \otimes {X_6}$ and
${Z_3} \otimes {Z_6}{X_6}$  to obtain the state ${\left| T
  \right\rangle _{36}}$; that is, in this case, the success probability is 1 when $m = 1$.

Clearly, Alice can achieve four possible results in $\left\{ {0,1,2,3}
\right\}$, and she will obtain one of the results for every
measurement with the same possibility. Thus, the possibility of
obtaining $m\left( {m = 0,1,2,3} \right)$ is 1/4. Similarly to when $m
= 1$, the possibility of obtaining one of the other three states is also 1. Therefore, the total success probability can be calculated as below
\begin{equation}\label{eq:schemeP}
{P_{suc}} = 4 \times \left( {\frac{1}{4} \times 1} \right) = 1.
\end{equation}

\subsection{Efficiency analysis}
\label{sec:3.2}

In RSP protocols, resource consumption, including quantum resources
and classical information, is an important criterion used to evaluate
a protocol. Taking five JRSP schemes for the preparation of
cluster-type states--ZHM11, ABD11, WY12, WY13 and H13--as a reference,
we perform analyses from the perspectives of quantum consumption,
classical information consumption, number of controlled-NOT gates and
success probability. For example, in the ZHM11 protocol, six Bell
states (12 qubits) are needed as quantum resources for the preparation of
a cluster-type state and 0 auxiliary qubits. In addition, the
measurement results of four qubits (8 bits of classical information)
are sent from the senders to the receiver, and no controlled-NOT gate
is involved in the process. Using the above calculation method, we can
obtain the efficiency information of the other four schemes (ABD11, WY12, WY13, H13) and our scheme (see Table~\ref{table2}).

\label{table2}
\begin{table}[h]\normalsize
\caption{
{\bf Comparison between our scheme and analogous schemes.}}
\renewcommand\arraystretch{1.5}\begin{tabular}{p{2cm}p{4cm}p{3.6cm}p{3.5cm}p{2cm}}
\hline {\bf scheme }& {\bf quantum consumption (qbit) }&{\bf classical information consumption (bit)}& {\bf controlled-NOT gates (number) } & {\bf success probability }\\
\hline
  ZHM11  & 12(12+0) & 8 & 0(0+0) & $< 1$ \\
  ABD11 & 8(6+2) & 4 & 2(0+2)	& $< 1$ \\
  WY12  & 9(8+1) & 4 & 6(6+0)	& $< 1$ \\
  WY13  & 10(6+4)	& 4	& 4(0+4) & $ < 1$ \\
  H13 & 9(6+3) & 4 & 4(2+2) &	$<1$ \\
  Our scheme & 8(6+2) & 6 & 2(0+2) & $=1$ \\
\hline
\end{tabular}
\end{table}

From the perspective of quantum consumption in Table 2, our scheme
only needs 8 qubits, equivalent to ABD11 and fewer than the
others. With respect to the classical information consumption, our
scheme sends 6 bits of classical information, fewer than ABD11, WY12,
WY13 and H13. Additionally, except for ZHM11, the time of \emph{CNOT}
operation is the same as ABD11 and far less than WY12, WY13 and
H13. Most importantly, the receiver can achieve unit success
probability using our proposed scheme, whereas the success
probabilities of the other schemes are less than 1. Therefore, our scheme is more economical and practicable.

\section*{Conclusion}
In this paper, we proposed a new scheme to realize the deterministic
joint remote preparation of a four-qubit cluster-type state utilizing
two GHZ states as quantum channels. In our scheme, the first sender
performs a two-qubit projective measurement based on the real
coefficient of the desired state. Then, the other sender performs
another projective measurement utilizing the measurement result and
the complex coefficient. To obtain the desired state, the receiver
applies appropriate unitary operations to his/her own two qubits and
two \emph{CNOT} operations to the two ancillary ones. The analysis
revealed that our scheme requires fewer qubit resources, less
classical information and no more \emph{CNOT} operations than most of
the analogous schemes. Furthermore, the receiver can achieve the
desired state with unit success probability. Hence, our scheme is somewhat more economical and feasible.

\begin{acknowledgements}
This work is supported by the National Nature Science Foundation of China (Grant Nos. 61502101, and 61501247), the Priority Academic Program Development of Jiangsu Higher Education Institutions (PAPD), the Natural Science Foundation of Jiangsu Province under Grant No. BK20140651.
\end{acknowledgements}



\end{document}